\documentstyle[twocolumn,aps,prl,floats,epsf]{revtex}

\begin{document}

\draft

\wideabs{

\title{$\bar d / \bar u$ Asymmetry and the Origin of the Nucleon Sea}

\author{
J.C.~Peng$^f$, 
G.T.~Garvey$^f$, 
T.C.~Awes$^i$,
M.E.~Beddo$^h$, 
M.L.~Brooks$^f$,
C.N.~Brown$^c$, 
J.D.~Bush$^a$,
T.A.~Carey$^f$, 
T.H.~Chang$^h$,
W.E.~Cooper$^c$,
C.A.~Gagliardi$^j$,
D.F.~Geesaman$^b$, 
E.A.~Hawker$^j$, 
X.C.~He$^d$,
L.D.~Isenhower$^a$,
S.B.~Kaufman$^b$, 
D.M.~Kaplan$^e$, 
P.N.~Kirk$^g$, 
D.D.~Koetke$^k$, 
G.~Kyle$^h$,
D.M.~Lee$^f$,
W.M.~Lee$^d$, 
M.J.~Leitch$^f$, 
N.~Makins$^b$\cite{byline1}, 
P.L.~McGaughey$^f$, 
J.M.~Moss$^f$,
B.A.~Mueller$^b$,
P.M.~Nord$^k$,
B.K.~Park$^f$, 
V.~Papavassiliou$^h$, 
G.~Petitt$^d$, 
P.E.~Reimer$^f$,
M.E.~Sadler$^a$,
J.~Selden$^h$, 
P.W.~Stankus$^i$, 
W.E.~Sondheim$^f$, 
T.N.~Thompson$^f$, 
R.S.~Towell$^a$\cite{byline2},
R.E.~Tribble$^j$,
M.A.~Vasiliev$^j$\cite{byline3}, 
Y.C.~Wang$^g$, 
Z.F.~Wang$^g$, 
J.C.~Webb$^h$, 
J.L.~Willis$^a$,
D.K.~Wise$^a$,
G.R.~Young$^i$\\ \vspace*{9pt}
(FNAL E866/NuSea Collaboration)\\ \vspace*{9pt}
}
\address{
$^a$Abilene Christian University, Abilene, TX 79699\\
$^b$Argonne National Laboratory, Argonne, IL 60439\\
$^c$Fermi National Accelerator Laboratory, Batavia, IL 60510\\
$^d$Georgia State University, Atlanta, GA 30303\\
$^e$Illinois Institute of Technology, Chicago, IL  60616\\
$^f$Los Alamos National Laboratory, Los Alamos, NM 87545\\
$^g$Louisiana State University, Baton Rouge, LA 70803\\
$^h$New Mexico State University, Las Cruces, NM, 88003\\
$^i$Oak Ridge National Laboratory, Oak Ridge, TN 37831\\
$^j$Texas A \& M University, College Station, TX 77843\\
$^k$Valparaiso University, Valparaiso, IN 46383
}
\date{\today}

\maketitle
\begin{abstract}
The Drell-Yan cross section ratios, $\sigma (p + d) / \sigma (p + p)$,
measured in Fermilab E866, have led to the first
determination of $\bar d(x) / \bar u(x)$, $\bar d(x) - \bar u(x)$, and
the integral of $\bar d(x) - \bar u(x)$ for the proton 
over the range $0.02 \le x \le 0.345$. 
The E866 results are compared with predictions based on 
parton distribution functions and various theoretical models.
The relationship between the E866 results and the NMC measurement 
of the Gottfried integral is discussed. The agreement between the
E866 results and models employing virtual mesons indicates these
non-perturbative processes play an important role in the origin 
of the $\bar d, \bar u$
asymmetry in the nucleon sea. 
\end{abstract} 
}

Recent measurements~\cite{nmc,na51,e866} have revealed a marked
asymmetry in the distributions of up and down quarks in the nucleon
sea. While no known symmetry requires $\bar u$ to equal $\bar d$, a
large $\bar d / \bar u$ asymmetry was not anticipated.  The
principal reason for expecting symmetry between up and down quarks in
the sea is an assumption that the sea originates primarily from
$q$-$\bar q$ pairs produced from gluons. As the masses of the up and
down quarks are small compared to the confinement scale, nearly equal
numbers of up and down pairs should result. Indeed, a theoretical
investigation~\cite{ross} of the light-quark asymmetry in the nucleon
concluded that perturbative processes do not give rise to asymmetries
in the up, down sea exceeding 1\%. Thus a large
$\bar d / \bar u$ asymmetry requires a non-perturbative origin for an
appreciable fraction of these light antiquarks.  This paper draws
together several implications arising from this observed $\bar d /
\bar u$ asymmetry -- the effect of these measurements on existing
parton distributions, an examination of the compatibility of the
measurements of this asymmetry, and the origin of the effect.

The issue of the equality of $\bar u$ and $\bar d$ was first
encountered in measurements of the Gottfried integral~\cite{gott},
defined as
\begin{equation}
I_G = \int_0^1 \left[F^p_2 (x,Q^2) - F^n_2 (x,Q^2)\right]/x~ dx,
\end{equation}
where $F^p_2$ and $F^n_2$ are the proton and neutron structure
functions measured in deep inelastic scattering (DIS)
experiments. 
$I_G$
can be expressed in terms of the valence and sea quark
distributions of the proton as:
\begin{eqnarray}
\lefteqn{I_G  = \frac{1}{3} \int_0^1 \left[u_v (x, Q^2) - d_v (x,Q^2)\right] dx} \hspace{0.9in} \nonumber \\
       & + \frac{2}{3} \int_0^1 \left[\bar u (x,Q^2) - \bar d (x,Q^2)\right] dx.
\label{eq:ig}
\end{eqnarray}
Under the assumption of a $\bar u$, $\bar d$ flavor-symmetric sea in
the nucleon, the 
Gottfried Sum Rule (GSR)~\cite{gott}, $I_G
= 1/3$, is obtained. Measurements of muon DIS on hydrogen and
deuterium by the New Muon Collaboration (NMC) 
determined that $\int
_{0.004}^{0.8} \left[F_2^p(x) - F_2^n(x)\right]/x~ dx = 0.221
\pm0.021$ at $Q^2 = 4$ GeV$^2$~\cite{nmc}.  Extrapolating to $x=0$
through the unmeasured small-$x$ region, the Gottfried integral is
projected to be $ 0.235\pm 0.026$, significantly below 1/3.

Although the violation of the GSR observed by NMC can be explained by
assuming pathological behavior of the parton distributions at
$x<0.004$, a more natural explanation is to abandon the assumption
$\bar u = \bar d$.  Specifically, the NMC result implies
\begin{equation}
\int_0^1 \left[\bar d(x) - \bar u(x)\right] dx = 0.148 \pm 0.039.
\end{equation}
The Fermilab E866 measurement~\cite{e866} of the ratio of Drell-Yan~\cite{dy}
yields from hydrogen and deuterium directly determines the ratio $\bar
d(x)/ \bar u(x)$ for $0.02<x<0.345$. An excess of $\bar d$ over $\bar
u$ is found over this $x$ range, supporting the observation
by NMC that the GSR is violated.

\begin{figure}
  \begin{center}
    \mbox{\epsffile{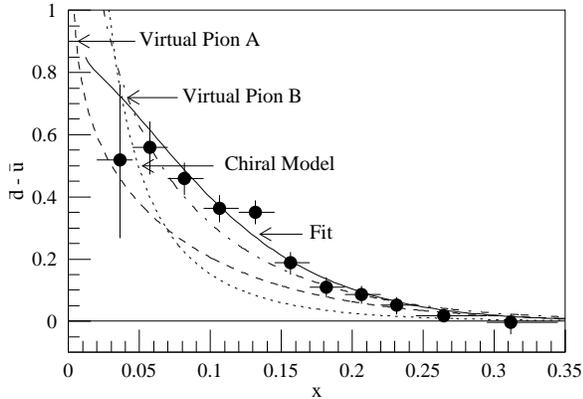}}
  \end{center}

  \caption{Comparison of the E866 $\bar d - \bar u$ results with the
  predictions of various models as described in the text.}

  \label{fig:dbar_fig1}
\end{figure}

The $\bar d / \bar u$ ratios measured in E866, together with the
CTEQ4M~\cite{cteq} values for $\bar d + \bar u$, were used to obtain
$\bar d - \bar u$ over the region $0.02 < x < 0.345$
(Fig.~\ref{fig:dbar_fig1}). As a flavor non-singlet quantity, $\bar
d(x) - \bar u(x)$ has the property that its integral is
$Q^2$-independent~\cite{martin}. Furthermore, it is a direct measure
of the contribution from non-perturbative processes, since
perturbative processes cannot cause a significant $\bar d$, $\bar u$
difference.  As shown in Fig. 1, the $x$ dependence of $\bar d - \bar
u$ at $Q$ = 7.35 GeV can be approximately parametrized as $0.05
x^{-0.5} (1-x)^{14} (1+100x)$.

Integrating $\bar d(x) - \bar u(x)$ from E866, one finds
\begin{eqnarray}
\lefteqn{\int_{0.02}^{0.345} \left[\bar d(x)  - \bar u(x)\right] dx = } \hspace{0.75in}\nonumber \\
       & 0.068\pm 0.007\text{(stat.)}\pm 0.008\text{(syst.)} 
\end{eqnarray}
at $Q = 7.35$~GeV~\cite{e866}.  To investigate the compatibility of
this result with the NMC measurement (Eq. 3), the contributions to the
integral from the regions $x < 0.02$ and $x > 0.345$ must be
estimated.  Table~\ref{tab:tab1} lists the values for the integral of
$\bar d - \bar u$ over the three regions of $x$ for three different
parton distribution function (PDF) parametrizations at $Q=7.35$
GeV. For $x >0.345$, the contribution to the integral is small (less
than 2\%).  The three parametrizations predict that the bulk of the
contribution to the integral comes from $0.02 < x <0.345$.  Since
CTEQ4M provides a reasonable description of the E866 data in the
low-$x$ region~\cite{e866}, and the contribution from the high-$x$
region is small, we have used CTEQ4M to estimate the contributions to
the integral from the unmeasured $x$ regions. This procedure results
in a value $\int_0^1 \left[\bar d(x) - \bar u(x)\right] dx = 0.100 \pm
0.007 \pm 0.017$, which is $2/3$ the value deduced by NMC. The
systematic error includes the uncertainty ($\pm 0.015$) due to the
unmeasured $x$ regions, estimated from the variation between CTEQ4M
and MRS(R2)~\cite{mrs}. This result is consistent with the integral of
the parametrized fit shown in Fig. 1.

\begin{table}

  \caption{Values for $\int \left[\bar d(x) - \bar u(x)\right] dx$ 
  over various $x$ ranges, evaluated at $Q$ = 7.35 GeV, for 
  various PDF parametrizations. Values deduced from E866 are also listed.}
  \label{tab:tab1}
  \begin{tabular}{ccccc}
$x$ range & CTEQ4M & MRS(R2) & GRV94~\cite{grv} & E866\\ \hline
0.345 - 1.0 & 0.00192 & 0.00137 & 0.00148 &\\ 
0.02 - 0.345 & 0.0765 & 0.1011 & 0.1027 & 0.068 $\pm$ 0.011\\ 
0.0 - 0.02 & 0.0296 & 0.0588 & 0.0584 &\\ 
0.0 - 1.0 & 0.1080 & 0.1612 & 0.1625 & 0.100 $\pm$ 0.018\\ 
  \end{tabular}
\end{table}

\begin{figure}
  \begin{center}
    \mbox{\epsffile{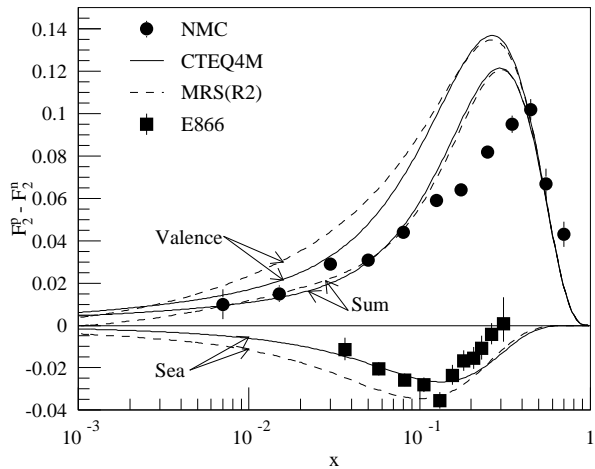}}
  \end{center}

  \caption{$F^p_2 - F^n_2$ as measured by NMC compared with predictions
  based on the CTEQ4M and MRS(R2) parametrizations. Also shown
  are the E866 results for the sea-quark contribution to
  $F^p_2 - F^n_2$. For each prediction, the top (bottom) curve is the
  valence (sea) contribution and the middle curve is the sum of the two.}

  \label{fig:dbar_fig2}
\end{figure}

The difference between the NMC and E866 results for the $\bar d - \bar
u$ integral raises the question of the compatibility of the two
measurements.  Figure~\ref{fig:dbar_fig2} shows the NMC data for
$F_2^p - F_2^n$ at $Q$ = 2 GeV, together with the fits of MRS(R2) and
CTEQ4M. Both PDF parametrizations give very similar results for $F_2^p
- F_2^n$. However, their agreement with the NMC data is poor,
especially in the region $0.15 < x < 0.4$.  It is instructive to
decompose $F_2^p(x) - F_2^n(x)$ into contributions from valence and
sea quarks:
\begin{eqnarray}
\lefteqn{F_2^p(x) -F_2^n(x) = }\hspace{0.75in} \nonumber \\
 & {1 \over 3} x \left[u_v(x) - d_v(x)\right] + {2 \over
3} x \left[\bar u(x) - \bar d(x)\right].
\end{eqnarray}
Two PDF parametrizations of these contributions are shown in
Fig.~\ref{fig:dbar_fig2}.  The valence contribution is positive, while
the contribution from the sea is negative. The MRS(R2) and CTEQ4M
parametrizations give noticeably different values for the valence and
sea contributions, though their net results for $F_2^p - F_2^n$ are
very similar. As shown in Fig.~\ref{fig:dbar_fig2}, the E866 data
provide a direct determination of the sea-quark contribution to $F_2^p
- F_2^n$, and can be used to distinguish between different PDF
parametrizations that produce similar fits to the NMC data. As the
direct determination of $\bar d(x) - \bar u(x)$ is smaller than
obtained from either PDF set, the parameters must be adjusted to
reduce the magnitude of both the sea and valence distributions in the
interval $0.03 \le x \le 0.3$.  This reduction will force an increase
in the valence contribution to the integral from $x \le 0.03$ and
could therefore bring the results from E866 and NMC into better
accord.  Fig.~\ref{fig:dbar_fig2} also suggests that the reason for
the difference between the PDF fits and the NMC results in the
interval $0.15<x<0.4$ is that the PDFs cannot accommodate the rapid
variation in the asymmetry of the nucleon sea as a function of $x$
revealed by E866.

The E866 data also allow the first determination of the momentum
fraction carried by the difference of $\bar d$ and $\bar u$.  We
obtain $\int_{0.02}^{0.345} x \left[\bar d(x) - \bar u(x)\right] dx =
0.0065 \pm 0.0010$ at $Q$ = 7.35 GeV.  If CTEQ4M is used to estimate
the contributions from the unmeasured $x$ regions, one finds that
$\int_0^1 x \left[\bar d(x) - \bar u(x)\right] dx = 0.0075 \pm
0.0011$, roughly 3/4 of the value obtained from the PDF
parametrizations.  Unlike the integral of $\bar d(x) -
\bar u(x)$, the momentum integral is $Q^2$-dependent and decreases as
$Q^2$ increases.

We now turn to the origin of the $\bar d / \bar u$ asymmetry.  As
early as 1982, Thomas~\cite{thomas} pointed out that the virtual pions
that dress the proton will lead to an enhancement of $\bar d$ relative
to $\bar u$ via the (non-perturbative) 
``Sullivan process.''
Sullivan~\cite{sullivan} previously showed that in DIS these
virtual mesons scale in the Bjorken limit and contribute to the
nucleon structure function.  Following the publication of the NMC
result, many papers [13-20] have treated virtual
mesons as the origin of the asymmetry in the up, down sea of the
nucleon.

Using the notion that the physical proton ($p$) may be expanded in a
sum of products of its virtual meson-baryon (MB) states, one writes $p
= (1-\alpha) p_0 + \alpha \text{MB}$, where $\alpha$ is the
probability of the proton being in virtual states MB and $p_0$ is a
proton configuration with a symmetric sea. It is easy to
show~\cite{kumano,szczurek} that
\begin{equation}
\int_0^1\left[\bar d(x,Q^2)-\bar u(x,Q^2)\right] dx = (2a - b) / 3
\end{equation}
where $a$ is the probability of the virtual state $\pi N$ and $b$ the
probability for $\pi \Delta$. These two configurations are the
dominant intermediate MB states contributing to the
asymmetry~\cite{szczurek,koepf}. Further, most recent calculations of
the relative probability of these two configurations find $a
\approx 2b$~\cite{szczurek,koepf}.  Using the value for the integral
extracted from E866 and assuming $a = 2b$ yields $a = 2b =
0.20\pm0.036$, requiring a substantial presence of virtual mesons in
the nucleon in this model. Following the observation~\cite{ehq} that
these configurations have a large impact on the spin structure of the
nucleon because pion emission induces spin flip, one can show that
\begin{equation}  
\Delta u_p - \Delta d_p =5/3 - 20(2a+b) / 27 = 1.296\pm 0.067
\end{equation}
using the above values of $a$ and $b$ determined from the E866
result. Here $\Delta u_p$ ($\Delta d_p$) is the total spin carried by
up (down) quarks in the proton.  This value is in good agreement with
the measured axial coupling constant for the nucleon,
$g_A=1.260\pm0.003$~\cite{pdb}, and increases confidence in the
virtual meson-baryon picture.

The $x$ dependences of $\bar d - \bar u$ and $\bar d / \bar u$
obtained in E866 provide important constraints for theoretical models.
Fig.~\ref{fig:dbar_fig1} compares $\bar d(x) - \bar u(x)$ from E866
with a virtual-pion model calculation following the procedure detailed
by Kumano~\cite{kumano}.  Since the E866 results are shown at a $Q$ of
7.35 GeV, the SMRS(P2)~\cite{smrs} parametrization for the pion
structure functions at this $Q$ is employed. The curve labeled
``virtual pion A" in Fig.~\ref{fig:dbar_fig1} uses a dipole form with
$\Lambda = 1.0$ GeV for the $\pi N N$ and $\pi N \Delta$ form factors,
and is seen to underpredict the magnitude of $\bar d - \bar
u$. However as has been noted~\cite{szczurek,koepf}, $\Delta$
production experiments~\cite{stoler} suggest a considerably softer
form factor for $\pi N \Delta$ than for $\pi N N$. Indeed much better
agreement with the E866 results is obtained by reducing $\Lambda$ for
the $\pi N \Delta$ form factor to 0.8 GeV, as shown by the curve
labeled ``virtual pion B" in Fig.~\ref{fig:dbar_fig1}. This fit
produces a value of 0.11 for the integral of $\bar d - \bar u$ and
1.18 for $g_A$. If $\Lambda$ is chosen to be 0.9 GeV (0.7 GeV) for the
$\pi N N$ ($\pi N \Delta$) form factor, one finds nearly exact accord
with the values cited in the previous paragraph.

A different approach for including the effects of virtual mesons has
been presented by Eichten, Hinchliffe, and Quigg~\cite{ehq} and
further investigated by Szczurek {\em et al.}~\cite{szczurek2}. In the
framework of chiral perturbation theory, the relevant degrees of
freedom are constituent quarks, gluons, and Goldstone bosons. In
this model, a portion of the sea comes from the couplings of Goldstone
bosons to the constituent quarks, such as $u \to d \pi^+$ and $d \to u
\pi^-$. The excess of $\bar d$ over $\bar u$ is then simply due to the
additional up valence quark in the proton.  The predicted
$\bar d - \bar u$ from the chiral model is shown in
Fig.~\ref{fig:dbar_fig1} as the dotted curve. We follow the
formulation of Szczurek {\em et al.}~\cite{szczurek2} to calculate
$\bar d(x) - \bar u(x)$ at $Q$ = 0.5 GeV, and then evolve the results
to $Q$ = 7.35 GeV. In the chiral model, the mean-$x$ of $\bar d -
\bar u$ is considerably lower than in the virtual-pion
model just considered. This difference reflects the fact
that the pions are softer in the chiral model, since they are coupled
to constituent quarks which on average carry only 1/3 of the
nucleon momentum. The $x$ dependence of the E866 data favors the
virtual-pion model over the chiral model, suggesting that
correlations between the chiral constituents should be taken into
account.

Another non-perturbative process that can produce a $\bar d$, $\bar u$
asymmetry is the coupling of instantons to the valence quarks. An
earlier publication~\cite{inst} presented an asymmetry due to
instantons but parametrized the result in terms of the asymmetry
observed in NMC, and therefore has no independent
predictive power.  Also the {\em ad hoc} $x$ dependence used for
$\bar d(x) /\bar u(x)$ is in poor agreement with the E866 result.
We are unable to determine if better agreement with data can be
obtained by an improved parametrization within the instanton model.

\begin{figure}
  \begin{center}
    \mbox{\epsffile{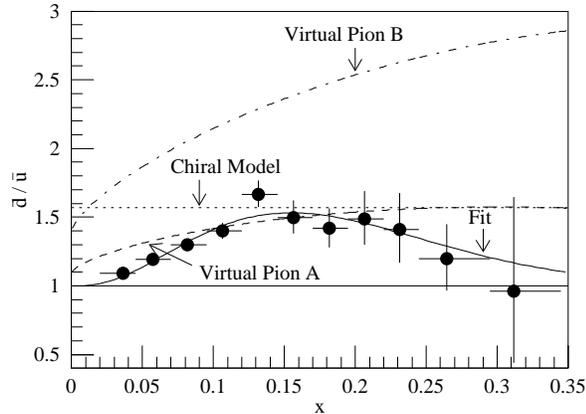}}
  \end{center}

  \caption{Comparison of E866 $\bar d / \bar u$ ratios at $Q$ =
  7.35 GeV with the predictions of two virtual pion models and the
  chiral model. The fit $1 + 1120 x^{2.75}
  (1-x)^{15}$ is also shown.}

  \label{fig:dbar_fig3}
\end{figure}

It is also instructive to compare the model predictions of $\bar d(x)
/ \bar u(x)$ with the E866 results. Figure~\ref{fig:dbar_fig3} shows
that the two virtual-pion models and the chiral model give $\bar d(x)
/ \bar u(x)$ values very different from the E866 result.  Note that
these calculations do not include the perturbative processes $g \to u
\bar u, d \bar d$ which generate a symmetric sea. Indeed, the $\bar d
/ \bar u$ data provide valuable information on the relative importance
of the perturbative (symmetric) versus the non-perturbative sea.  The
chiral model predicts a $\bar d / \bar u$ ratio of 11/7 for all $x$.
Figure~\ref{fig:dbar_fig3} shows that the observed ratio nearly equals
this value for $0.1 < x < 0.2$, leaving little room for any
perturbatively generated symmetric sea in this interval, which seems
unreasonable. The same problem also arises for the virtual-pion model
A. In contrast, the virtual-pion model B readily accommodates
contributions from a symmetric sea.

In summary, E866 has provided the first determination of $\bar d /
\bar u$, $\bar d - \bar u$, and the integral of $\bar d - \bar u$ over
the range $0.02 \le x \le 0.345$.  It provides an independent
confirmation of the violation of the Gottfried Sum Rule reported from
DIS experiments. The magnitude of the integral of $\bar d - \bar u$
over the region $0.02 < x < 0.345$ is smaller than obtained from 
some current PDF parametrizations. This indicates that the violation of
the Gottfried Sum Rule is likely smaller than reported by
NMC. Together with the NMC data, the E866 results impose stringent
constraints on both sea- and valence-quark distributions. The good
agreement between the E866 $\bar d -
\bar u$ data and the virtual-pion model indicates that virtual
meson-baryon components play an important role in determining
non-singlet structure functions of the nucleon. Future experiments
extending the measurements of $\bar d / \bar u$ to other $x$ and
$Q^2$ regions can further illuminate the interplay between the
perturbative and non-perturbative elements of the nucleon sea.

This work was supported in part by the U.S. Department of Energy.

\end{document}